# Modelling air pollution abatement in deep street canyons by means of air scrubbers

Marina De Giovanni[1,*], Gabriele Curci[2,3], Alessandro Avveduto[1], Lorenzo Pace[1], Cesare Dari Salisburgo[1], Franco Giammaria[2,3], Alessio Monaco[3], Giuseppe Spanto[4], Paolo Tripodi[1,4]

[1] Società Progetti Innovativi – S.P.In Srl – Via Carlo D'Andrea snc, 67100 Bazzano, L'Aquila

[2] CETEMPS, Centre of Excellence for the integration of remote sensing techniques and numerical modelling for the forecast of severe weather, University of L'Aquila, L'Aquila, Italy

[3] Dept. of Physical and Chemical Sciences, University of L'Aquila, L'Aquila, Italy

[4] Innovation in Sciences & Technologies – IsTECH Srl – Via Mar della Cina 304, 00144 Roma

* Corresponding author: marina.degiovanni@spintecnologia.com

# Abstract

Deep street canyons are characterized by weak ventilation and recirculation of air. In such environment, the exposure to particulate matter and other air pollutants is enhanced, with a consequent worsening of both safety and health. The main solution adopted by the international community is aimed at the reduction of the emissions. In this theoretical study, we test a new solution: the removal of air pollutants close to their sources by a network of Air Pollution Abatement (APA) devices. The APA technology depletes gaseous and particulate air pollutants by a portable and low-consuming scrubbing system, that mimics the processes of wet and dry deposition. We estimate the potential pollutant abatement efficacy of a single absorber by Computational Fluid Dynamics (CFD) method. The presence of the scrubber effectively creates an additional sink at the bottom of the canyon, accelerating its cleaning process by up to 70%, when an almost perfect scrubber (90% efficiency) is simulated. The efficacy of absorber is not proportional to its internal abatement efficiency, but it increases rapidly at low efficiencies, then tends to saturate. In the particular configuration of the canyon we choose (aspect ratio of 3) the upwind corner is the most favourable for the absorber application. In the downwind corner the maximum pollutant abatement is -24%, while in the upwind corner the maximum abatement is twice as much at -51%. The efficacy of the absorber increases much faster at low efficiencies: at 25% efficiency, the abatement is already about half that obtained with the 90% efficiency. These first results, suggest strategies for the real-world application of a network of absorbers, and motivates further theoretical study to better characterize the details of the air flow inside and around the absorber.

# 1. Introduction

A growing body of scientific evidence suggests that air pollution is more harmful than previously thought (European Environmental Agency, EEA, 2013; Beelen et al., 2013). This was reinforced by a statement in October from the International Agency for Research on Cancer (IARC 2013), which classified air pollution as carcinogenic. The risk for human health depends on both exposure concentration level and length. Long term exposure to airborne particulate matter (PM) has been associated to increased cardiovascular mortality, to various blood markers of cardiovascular risk, to histopathological of subclinical chronic inflammatory lung injury and subclinical atherosclerosis, while short term exposure to PM has been associated to cardiovascular mortality and hospital admission (Pope and Dockery, 2006; Beelen et al., 2013). Unfortunately, the problem of air pollution is far from being solved. In a recent study, EEA showed that, between 2009 and 2011, 96% of city dwellers were exposed to fine particulate matter (PM2.5) concentrations above the World Health Organization (WHO) guidelines and 98% were exposed to ozone ($O_3$) levels above WHO guidelines (EEA, 2013). The estimated number of deaths cause worldwide caused by air pollution is around 7 million people (http://www.who.int/mediacentre/news/releases/2014/air-pollution/en/).

In view of potentially large benefit associated with abatement of air pollution in term of health and costs, the main solution adopted by the international community is aimed at the reduction of the emissions. We are experimenting a new alternative approach: removal air pollutants close to their sources by a mean network of air pollution Air Pollution Abatement (APA) devices. The APA technology depletes air pollutant (particulate matter, heavy metals, Polycyclic Aromatic Hydrocarbons, Light Hydrocarbons, $NO_x, SO_x, CO_2$) by an innovative system that mimics the processes of wet and dry deposition. The air cleaner system employed in this study is a prototype of a small three-stage wet scrubber characterized by low consumption, easily operating and adaptable to be installed in network in confined and urban areas. Currently, the scrubber processes about 2300 $m^3$/h of ambient air, requiring a power of 550 W and producing a noise of 55 dBA/1m. The first stage is composed by a weak Venturi scrubber system with an appropriate water sprayer. The second stage is performing a shower scrubbing. Finally, the third stage is a variable deposition stack where the air flow is forced through a series of disks with properly arranged holes, in order to force deposition onto the disk surfaces.

There are many potentially different favourable pseudo-indoor locations for the installation of the absorbers: metropolitan, underground car parks, galleries etc.. Here we focus our attention on the case of the deep street canyon. The street canyon is an urban street surrounded by tall buildings. This peculiar urban habitat traps the pollutants and limits their dispersal into the atmosphere and people who lives or works in a street canyon may be exposed to high concentrations of pollutants for a long time.

We briefly review and summarize what is known on the wind flow patterns inside the canyon and on the pollutant's concentrations and their spatial distribution. A fundamental parameter in determining the field of wind flow within the canyon is the aspect ratio H/W, where H is height of the buildings and W is the width of the street. For a wind perpendicular to the street axis, Oke (1988) distinguished three flow regimes depending on the value H/W:

- H/W < 0.3 Isolated Roughness Flow (IRF). The distance between the buildings is such as to ensure that the field of wind flow between the two buildings is unruffled.
- 0.3 < H/W < 0.67 Wake Interference Flow (WIF). The distance between the buildings is such as to ensure that the field of wind flow between the two buildings is wake unruffled.
- H/W > 0.67 Skimming Flow (SF). In this case it was observed the presence of vortices and the wind flow pattern in the canyon is separated from the air stream above (Xian, 2006).

The SF regime is characterized by low ventilation and recirculation of air, therefore the pollutants emitted in the canyon stagnate there. In this work we focus on the skimming flow case. The number of vortices generated inside the canyon strongly depends on the value of H/W. For H/W=1 a vortex is formed (Hunter 1992), for H/W=3 two vortices are formed, and for H/W=5 we have three vortices. The strength of the vortices depends mainly on the speed of the wind blowing over the roofs and is roughly directly proportional (Vardoulakis et al., 2003). When more vortices are present, their intensity increases with height from the street floor (Murena et al., 2011). The pollutants' concentration decrease with height and accumulates in the leeward side (Vardoulakis et al., 2003). When H/W $\geq$ 2 then the street canyon is classified as deep, and the pollutant's concentration is usually enhanced (Assimakoupolos et al., 2003).

The simplest model of street canyon is a flat street surrounded by tall buildings of the same constant height. However, in the real world the buildings don't have the same height and the canyon is asymmetric. There are two possible configurations: the step-up canyon, where the upwind building is lower than downwind building, and the step-down canyon, where the upwind building is higher than downwind building. The canyon's asymmetry influence the pollution dispersion pattern, and the concentration of pollutants is expected to increase in the step-down configuration (Assimakoupolos et al., 2003).

Another crucial factor in determining the spatial and temporal distribution of pollutants is the differential heating of the canyon internal surfaces. For a canyon with a ratio H/W=1.2, Sini et al. (1996) found that the heating of the leeward wall determines an increase in the intensity of the vortex and consequently an increase in the speed of air exchange between the bottom of the road and the atmosphere above the canyon.

Here we employ a Computational Fluid Dynamics (CFD) method to study the spatial and temporal distribution of a passive tracer within a deep street canyon in the presence of an absorber of pollution at street level. As detailed in section 2, we used the ANSYS Fluent software to simulate a 2D idealized deep street canyon with an aspect ratio H/W = 3, and we simulate the effect of the absorber as region of space where a depleting chemical reaction for the passive tracer occur. Results of simulations are illustrated in section 3, where we first describe the simulated wind flow pattern calculate the cleansing time scale of the canyon without and with the absorber. We then evaluate the abatement efficacy of the absorber, simulating a constant emissions of the passive tracer at street level and the presence of an absorber at several efficiency rates. We summarize our results in final section 4.

## 2. Methods

We employ the commercial software ANSYS Fluent to simulate the dynamics into a 2-D deep street canyon section, following Murena et al. (2011). Figure 1 shows the computational domain. The height H of the canyon is 18 m and its width W is 6 m for an aspect ratio H/W = 3. The domain sizes are such that the flow is not affected by the presence of the canyon and by the boundary conditions (Sini et al., 1996; Solazzo and Britter, 2007). The domain is dived into 50,000 grid points. We adopted the segregated solver and pressure-velocity coupling was carried out using the SIMPLE algorithm (Patankar and Spalding 1972). The κ-ε turbulence model was chosen. The roofs, buildings and street are set on wall-boundary with a no-slip condition. The right side and the top of the domain are set to pressure-outlet. The domain left border is a velocity inlet, with constant wind speed and direction perpendicular to the canyon axis. In the reference run, the wind speed at inlet is set to 2 m/s.

We carry out two types of simulations, where the generic pollutant concentration is represented by a passive tracer of the same molar mass of carbon monoxide. The first type of simulation is aimed at evaluating the cleansing time of the canyon. The simulation have zero boundary condition and starts with a uniform initial condition of the tracer. The average concentration of the tracer is monitored inside the canyon as a function of time, and the cleansing time is calculated as the time required to reach half of the initial concentration, following the analysis of Murena et al. (2011).

The second type of simulation is aimed at evaluating the pollutant abatement efficacy of the absorber in a realistic case. In this case, a constant emissions of the tracer from the street floor is simulated (to mimic traffic emissions) and a constant concentration boundary condition at the velocity inlet is imposed, in order to simulate the contribution of the urban background to the canyon tracer concentration.

The presence of the pollutant absorber APA is not simulated in detail, but with a simplified approach described as follows. As mentioned in the introduction, APA is able to process 2300 m$^3$/h of air. Considering the physical dimensions of the absorber (about 1.5 m width and 2 m height) and assuming a radius of influence of a few meters of the inlet and outlet flows, we represent the absorber as an area in the canyon section of 2 m width and 3 m height. This means that the simulated section is representative of a volume of air given by the assumed area of the absorber by a depth that depends on the speed of the flow passing through it and that is constrained by the nominal internal flow of 2300 m$^3$/h. For example, considering a typical wind speed of 0.025 m/s near the bottom of the canyon, with direction approximately parallel to the street floor, we calculate that the air flows through the width of 2 m of the absorber 45 times per hour. The depth representative of the APA effect is given by 2300 m$^3$/h divided by (2 × 3 m$^2$ × 45 h$^{-1}$), yielding 8.5 m. In that case, the results obtained in the 2-D simulation may be considered extended to a depth of 8.5 m along the canyon axis. This gives a mean to interpret the results in view of an hypothetical application of a series of absorbers along a street canyon.

The abatement of the pollutant is simulated via a loss reaction for the tracer:

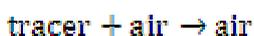

$$\text{tracer} + \text{air} \rightarrow \text{air}$$

which is converted to air at a constant rate. The reaction takes place in the 2 × 3 m$^2$ are represented by the absorber, and the rate is set as to obtain a desired abatement efficiency. In our simulations, the APA efficiencies simulated are 15%, 25%, 70%, and 90%. The simulations are repeated selecting the position of the APA alternatively at the upwind and downwind bottom corners of the street canyon.

## 3. Results

Figure 2 shows the velocity streamlines at steady-state within the urban canyon. As expected, for a canyon with H/W = 3 we see the presence of two vortices of opposed circulations, with the vortex on top having a strength greater than that below. The vortices are slowed down by friction, thus velocity decrease toward the bottom of the canyon. The lower vortex is more isolated and decoupled from the overlying wind structure, therefore the two vortices are expected to be characterized by different concentrations of pollutants. Moreover, for a canyon with H/W = 3 and a boundary wind speed of 2 m/s, the intensity of the vortex should be enough to make the concentration of passive pollutant uniform within each vortex (Murena et al., 2011).

Our first aim is to determine the halving-time of the average tracer concentration within the canyon with and without APA, in order to evaluate the modification to the street canyon cleansing capacity. The initial condition is a mixture of air and a passive tracer, assumed to be carbon monoxide (CO), at constant molar mixing ratio of $5 \times 10^{-6}$ (5 ppm) throughout the domain. When the simulation starts, from the left side of the domain enters pure air (no CO), with at velocity 2 m/s. The wind induce the vortices in the canyon, that limit the dispersion of the pollutant in the free atmosphere. Above roofs clean air enters on the canyon upwind side, while dirty air exits the canyon form the downwind side.

Figure 3 shows the trend of the spatially averages concentration of pollutant with a APA placed in the downwind corner of the canyon. The black straight line identifies the concentration corresponding to 50% of the initial value: its intersections with the curves allow to estimate the time necessary to halve the concentration of the pollutant, which we use as a measure of the street canyon cleansing time (Murena et al., 2011). The canyon-average tracer concentration without the scrubber is halved after 1000 seconds. When a scrubber with an efficiency of 15% is simulated into the canyon, the cleansing time decreases to 820 s, with a decrease of -18%. With scrubber efficiencies of 25%, 70% and 90% the cleaning time decreased by -25%, -45%, and -55%, respectively.

The results for the APA in the upwind corner is shown in Figure 4. In this case, the average tracer concentration decreases much faster than the downwind case, because the vortex circulation favour the accumulation of the tracer on that side. Remarkably, at 15% APA efficiency, the reduction of tracer half-time is -58%, a reduction greater than that obtained with the 90% efficient APA on the downwind corner. The maximum reduction obtained with the maximum APA efficiency is of -70%.

The reason for the very different efficacy of the absorber positioned in the downwind and the upwind side of the canyon, may be better understood looking at the snapshot of the simulated tracer concentration of Figure 5. Each vortex is characterized by a different average concentration, with the concentration being higher in the vortex centre and on the leeward side in the upper vortex and in the windward side in the lower vortex. The raise of the concentration of pollutant in a corner is determined by the circulation of that vortex. The lower vortex conveys the pollutant in the right corner, so the absorber efficacy is improved, because higher concentrations are processed and thus abated.

In Table 1 we summarize the half times of the tracer and the related abatement efficacies. The presence of the scrubber effectively creates an additional sink at the bottom of the canyon accelerating the cleaning process of the canyon. The efficacy of absorber is not proportional to its internal abatement efficiency, but it increases rapidly at low efficiencies, then tends to saturate. This means that a significant acceleration of the canyon cleansing process might be obtained already with low abatement efficiencies, and that further and costly improvements of the efficiency might not be needed.

We should now evaluate the pollutant abatement efficacy of the absorber in a realistic case. We include a source term that may represent the traffic emissions near the street floor. The emission is simulated as a constant mass flow of 1 kg/s of a mixture of CO with a mass fraction of $1 \times 10^{-5}$ (10 ppm) and air coming out from the street floor space not overlapping with the APA absorber area. In addition, we impose a constant urban background concentration of the tracer of $1 \times 10^{-7}$ (0.1 ppm).

In Figure 5 and Figure 6 we show the time series of the canyon-average tracer concentration without and with APA at different efficiency, respectively with the absorber positioned in the downwind and upwind corner. In Table 2 we report the calculated efficacy of the scrubber after 2000 seconds of simulation. We see that the concentrations with respect to the reference case without the absorber are gradually reduced until they reach approximately a constant difference with the reference case after 1000 seconds. As also found with the experiments on cleansing time, the upwind corner is the most favourable for the absorber application. In the downwind corner the maximum abatement, obtained with the more efficient APA 90%, is -24%. In the upwind corner the maximum abatement is twice as much at -51%. The efficacy of the absorber increases much faster at low efficiencies: at 25% efficiency, the abatement is already about half that obtained with the 90% efficiency.

## 4. Conclusions

In this work we theoretically tested the applicability of a network of pollution absorbers in urban environments. We focused on the particular case of the deep street canyon, which is characterized by low ventilation and frequent recirculation of air. These conditions are favourable for pollutant accumulation, but at the same time allow to re-process several times the same air by means of a scrubbing system.

We first estimated the canyon cleaning time scaling due to the operation of an ideal absorber in a Computational Fluid Dynamics (CFD) framework. By imposing a constant initial concentration, we estimated the half-time, the time necessary to reduce the concentration to 50% of its initial value, without and with the presence of an air scrubber (APA). The presence of the scrubber effectively creates an additional sink at the bottom of the canyon accelerating the cleaning process of the canyon. We found that the maximum reduction of the canyon cleansing time, obtained with an APA efficiency of 90%, is -55% and -70% for an absorber positioned in the downwind and in the upwind side of the canyon, respectively. This difference is explicable taking into account the spatial distribution of the tracer within the canyon, induced by the vortex near the street floor. The efficacy of the absorber is not proportional to its internal abatement efficiency, but it increases rapidly at low efficiencies, then tends to saturate. This means that a significant acceleration of the canyon cleansing process might be obtained already with low abatement efficiencies, and that further and costly improvements of the efficiency might not be needed.

We then evaluated the pollutant abatement efficacy of the absorber in a case with a constant emissions of the tracer from the street floor (such as traffic emissions) and a constant urban background concentration. As also found with the experiments on cleansing time, the upwind corner is the most favourable for the absorber application. In the downwind corner the maximum abatement, obtained with the more efficient APA 90%, is -24%. In the upwind corner the maximum abatement is twice as much at -51%. The efficacy of the absorber increases much faster at low efficiencies: at 25% efficiency, the abatement is already about half that obtained with the 90% efficiency.

The results shown here suggest a strategy for the deployment of a network of absorbers in a street canyon. First, they must be placed on both sides of the street, because the abatement efficacy is increased in the upwind side. If above the canyon there is prevalent wind direction, which defines the up- and down-wind sides, more absorbers should be placed in the most probable upwind side. The distance of the absorbers must be consistent with the expected wind speed along the canyon axis. When the above-roofs wind direction is not perfectly perpendicular to the street axis, the perpendicular component may still trigger the formation of the vortices, but the component parallel to the street axis will advect the structure, basically forming a spiral-like motion. The distance of the absorbers must then be sufficient to allow efficient air processing in relation of the expected wind speed component parallel to the street axis.

Future work will be aimed at confirming and better characterizing the results presented here, through a more detailed simulation of (1) the absorber flow dynamics (e.g. through the simulation of a realistic fan system), (2) other source of turbulence (e.g. traffic), and (3) the interaction of the canyon flow with the air flow modified by the absorber.

**Tables and Figures**

*Table 1. Simulated cleansing time of the canyon without and with the absorber (APA). The percentage in case description refers to the simulated APA abatement efficiency. UP and DOWN denote the position of the APA in the upwind and downwind side of the canyon, respectively. The cleansing time is calculated the as the time needed to half the canyon-average initial concentration of the passive tracer.*

| Case | Cleansing time (s) | % Change with respect to No APA |
|---|---|---|
| No APA | 1000 | - |
| APA 15% DOWN | 820 | -18% |
| APA 25% DOWN | 750 | -25% |
| APA 70% DOWN | 550 | -45% |
| APA 90% DOWN | 450 | -55% |
| APA 15% UP | 420 | -58% |
| APA 25% UP | 400 | -60% |
| APA 70% UP | 330 | -67% |
| APA 90% UP | 300 | -70% |

*Table 2. Percent change of passive tracer average concentration in the canyon with respect to the reference case without the absorber. Simulated values are relative to concentrations after 2000 seconds from the start. Case are the same listed in Table 1.*

| Case | % Change with respect to No APA |
|---|---|
| APA 15% DOWN | -7% |
| APA 25% DOWN | -11% |
| APA 70% DOWN | -20% |
| APA 90% DOWN | -24% |
| APA 15% UP | -20% |
| APA 25% UP | -27% |
| APA 70% UP | -45% |
| APA 90% UP | -51% |

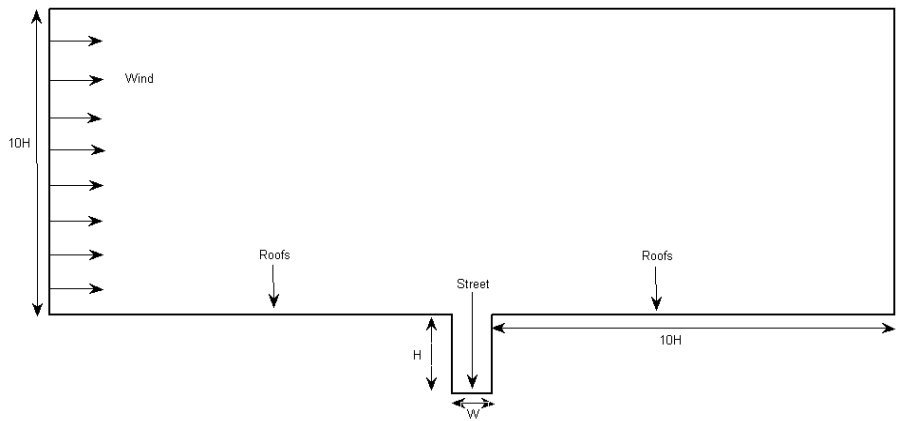

*Figure 1. Computational domain. In out simulations W = 6 m and H = 18 m, for an aspect ratio H/W = 3.*

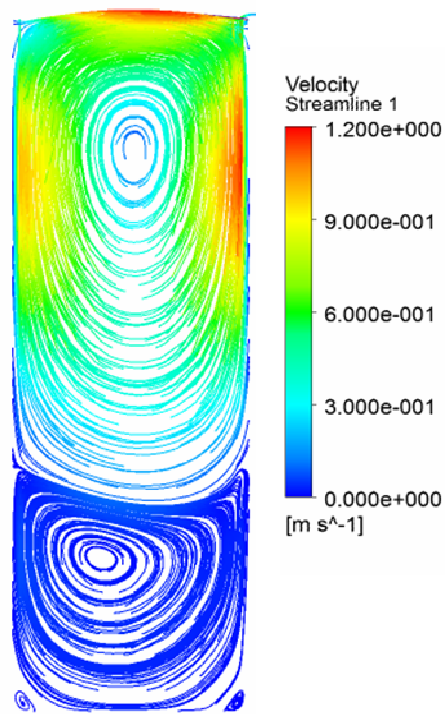

*Figure 2. Streamline of velocity calculated at steady-state in the street canyon*

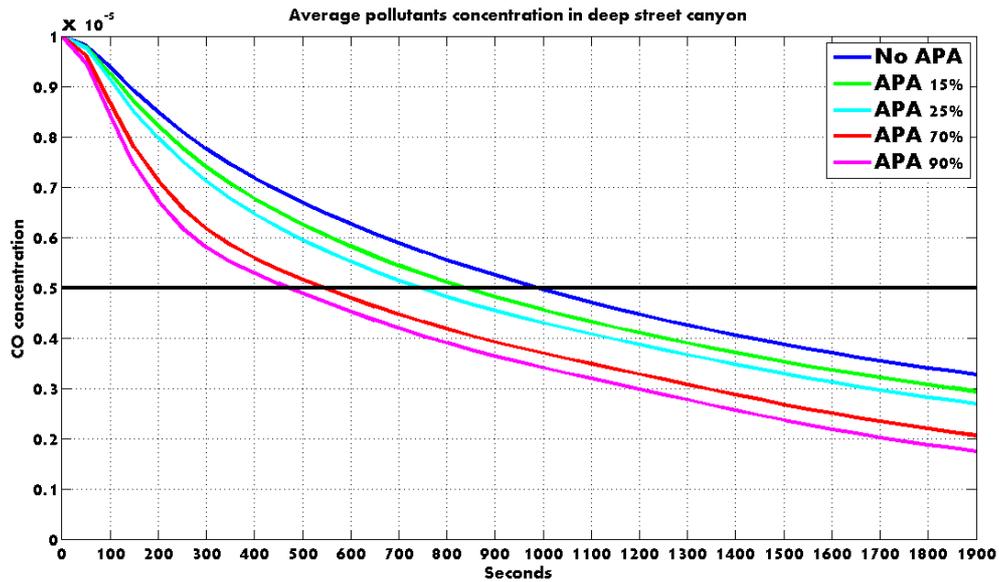

*Figure 3. Timeseries of average concentration of the passive tracer in the canyon with the absorber (APA) in the downwind side. Results are shown in the absence of the absorber and with the absorber at several abatement efficiencies. The black line denotes the half initial concentration value used to calculate the canyon cleansing time. Unit is tracer molar mixing ratio.*

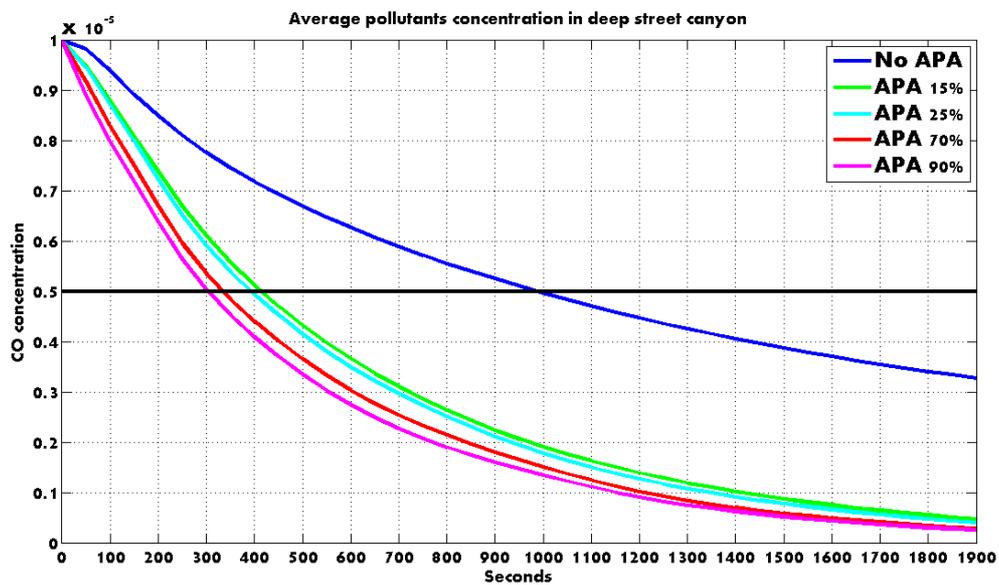

*Figure 4. Same as Figure 3, but for the pollutant absorber (APA) in the upwind side of the canyon.*

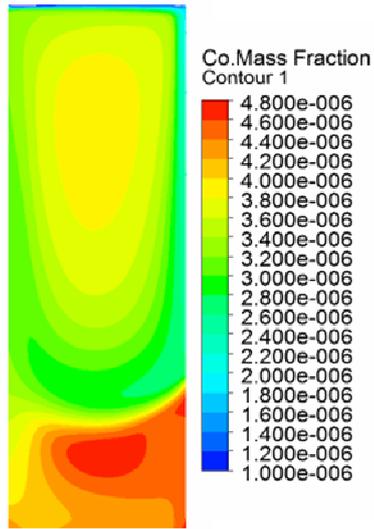

*Figure 5. Passive tracer concentration in the street canyon The figure is a snapshot of the simulation after 1300seconds from the start. Unit is tracer molar mixing ratio.*

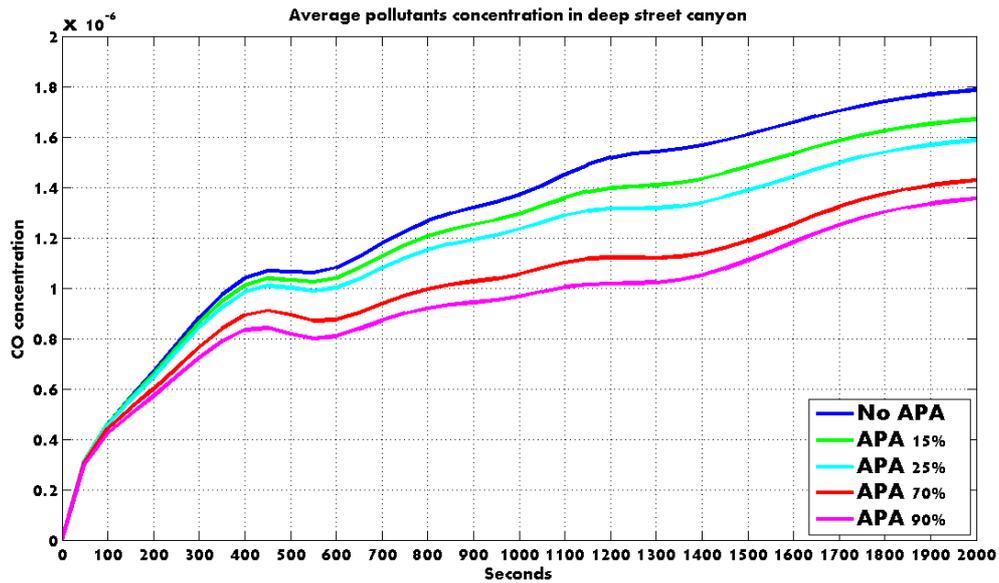

*Figure 6. Average passive tracer concentration in the street canyon, simulated in the presence of a constant emission from the street floor. Cases of APA in the downwind side of the canyon.*

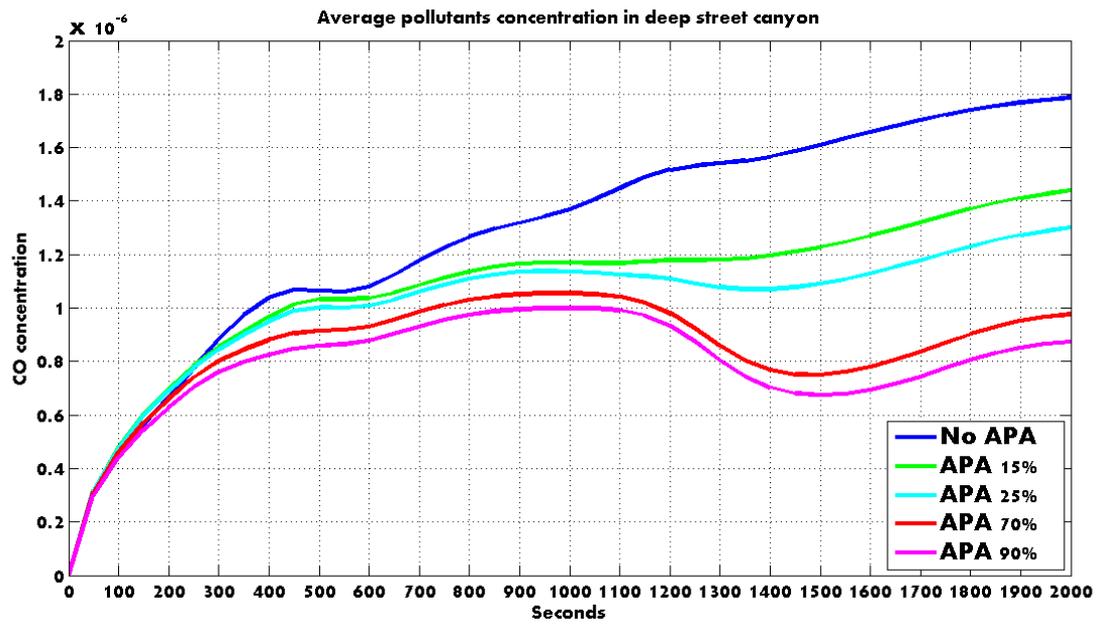

*Figure 7. Same as Figure 6, but for cases with APA in the upwind side of the canyon.*